\documentclass[aps,prl,twocolumn,10 pt,groupaddress,floats,showpacs]{revtex4}
\usepackage[dvips]{graphicx}
\begin{document}

\title{Comment to the paper : Collapse of the vortex-lattice inductance and shear
modulus at the melting transition in untwinned YBa$_2$Cu$_3$O$_7$, by Matl
\emph{et al.}}
\author{A.Pautrat, C. Goupil, Ch. Simon}
\address{CRISMAT/ISMRA, UMR 6508 associ\'ee au CNRS, Bd Mar\'echal Juin, 10450 Caen c\'edex, France}

\author{B. Pla\c{c}ais, P. Mathieu}
\address{LPMC de l'ENS, UMR-CNRS 8551 associ\'ee aux Universit\'es
Paris 6 et 7,  75231 Paris Cedex 5, France}

\date{\today}

\maketitle
 In a recent paper, Matl {et al.} (MOGT) present a high-frequency study of the
complex resistivity of a pinned vortex lattice in YBaCuO \cite{Matl}. They focus on the
inductive-to-resistive transition which is investigated as a function of temperature at a
constant field $B_0=2$ T, so that the transition is associated with the vanishing of
vortex pinning strength. The case of high-temperature materials is of particular interest
since the transition temperature $T_m$ is smaller by a few degrees than the diamagnetic
transition $T_c$. The nature of the inductive transition is still debated, with however,
strong indications from thermodynamics in favor of a first-order scenario. In this
context, the linear response is a very rich and useful probe that allows to characterize
both the resistive and inductive responses of the vortex state. As discussed in the
literature \cite {nono}, the high-frequency domain, including the pinning frequency
$\omega _p\lesssim 1$MHz in the soft materials such as YBaCuO crystals, is especially
relevant for elucidating the nature of the pinning mechanism. However, as drawback, it
requires to handle (tackle) experimental complications due to skin-effects.

The most salient feature of the MOGT-data is a strong divergence, followed
by a sharp discontinuity, of the low-frequency inductivity $L_s(T)$ at $T_m$
which is identified as the temperature of the first order transition. As it
is clear from the title of their paper, Matl {et al.} attribute this
discontinuity to the collapse of the vortex-lattice (VL) shear modulus $%
C_{66}$. We stress here that the observed, $\lambda $-shaped, temperature
dependence $L_s(T)$ is quite standard for a superconducting transition, and
rather indicative of a second-order transition; it can therefore not be
regarded as a proof of a collapse of $C_{66}$, even though this collapse is
very much reclaimed if the transition is a genuine melting.

As a main point, we recall that the relevant parameter for a superconducting
transition is not $L_s(T)$ but the effective superconducting-electron
density $\tilde{n}_s(T)$ which is inversely proportional to $L_s$. In the
vortex state the apparent superconducting density $\tilde{n}_s$ is
determined by the elastic pinning properties of the VL. Although $\tilde{n}_s
$ can be in principle deduced from the critical-current density, but it is
best and most directly related to the kinetic inductivity since in the
expression $L_s=m/(\tilde{n}_se^2)$ intervene only the elementary mass $m$
and charge $e$. Now, according to the experimental low-frequency data by
Matl et al. (see the 1 MHz data in Fig. 7 of \cite{Matl}), $1/L_s(T)\sim
(T_m-T)$ is linearly vanishing at $T_m$ which is the prescription for a
second-order transition. In order to understand to which part of the
experimental data, the shear discontinuity reported in the fig. 7 of Ref.%
\cite{Matl} can be traced back, we need to look into the details of the MOGT
data analysis.

If not related to the quasistatic response, the discontinuity must be due to
the finite-frequency deviations in the complex resistivity (dispersion). We
agree with the authors that the conventional Gittleman-Rosemblum model,
based on a purely local force-balance model with a single a return spring
constant $\kappa $ and a viscous drag coefficient $\eta $ \cite{nono,GR},
cannot explain the pinning frequency spectrum. In this respect, the proper
account of non-local effects is certainly the key ingredient for tackling
this long-standing disagreement. But this concern is very general and not
specific to the MOGT-data. As a matter of fact, MOGT-data are quite
representative of clean-YBaCuO spectra and very similar to what was
previously published \cite{pautrat} using a slightly different technics;the
YBaCuO-spectra themselves are not different from those of standard pinned
vortex states in low Tc materials \cite{nono}. In these materials, the
question of the depinning spectrum has been successfully solved by relying
on a simple model: the two-mode electrodynamics \cite{nono}. This model is
very well suited for pinning of vortices by surface irregularities, a
mechanism which is always present and prevails in clean materials, like
untwinned YBaCuO \cite{pautrat}. In this model, the shear modulus is not
relevant at all for dispersion but may intervene indirectly in the
phenomelogical parameter for the quasistatic response.

The situation with the MOGT analysis is quite opposite. Matl et al. use a
different model which relies on the existence of a dilute concentration of
very strong pinning centers with a matching field $B_\phi $ typically 100
times smaller than their working field $B_0=2$T. A further strong assumption
is that VL dynamics is purely two-dimensional. This hypothesis would need to
be controlled experimentally. This model is quite complicated; it contains,
beside that the onset frequency for skin-effect, three characteristic
frequencies: the pinning frequency $\omega _p=\kappa /\eta $ for the
strongly pinned vortex fraction (in the GHz range), the effective pinning
frequency $\tilde{\omega}_p=\omega _pB_\phi /B_0$ (in the 10 MHz range)
associated with the averaged pinning force $\tilde{\kappa}=\kappa B_\phi /B_0
$, and finally the shear frequency $\omega _{66}=4\pi C_{66}/\eta $ (in the
sub MHz range). Importantly for the analysis is the hierarchy $\omega _p\gg
\tilde{\omega}_p\gg \omega _{66}$, so that the low-frequency dispersion is
dominated by shear interactions. Note also that the quasistatic inductance
is given, up to a small logarithmic correction, by the classical GR formula
taking $\tilde{\kappa}$ as an effective spring constant, and therefore from
principle insensitive to $C_{66}$. Finally the reality of the three-decades
jump in $C_{66}$ which in their analysis precedes the true depinning
transition ($\kappa =0$) is very much dependent on the details of the
fitting procedure for the few frequency spectra taken in a the tiny
temperature range below $T_m$. To our view, the MOGT-conclusions rely on a
rather brittle experimental body and the collapse of C$_{66}$results from an
involved analysis of the finite frequency corrections to $\rho (\omega )$.
These corrections are not necessary since the complex frequency spectrum has
been previously interpreted by the two modes model, first proposed for low
Tc materials\cite{nono}. We think that it is more adequate to interpret the
present data and should be at least considered.

\end{document}